\documentclass[12pt]{article}
\usepackage{amssymb, amsmath, amsthm, amscd, graphicx}
\usepackage[all]{xy}

\begin{document}

\title{Interface Free Energy or Surface Tension:\\
definition and basic properties}

\author{
C.-E. Pfister\\
EPF-L, Institut d'analyse et calcul scientifique\\
B\^atiment MA, Station 8 \\
CH-1015 Lausanne, Switzerland\\
e-mail: charles.pfister@epfl.ch}

\date{November 2009}
\maketitle

\begin{abstract}
{\bf Interface free energy} is the contribution to the  free energy of a system  due to the presence of an interface separating two coexisting phases at equilibrium. It is also called interfacial free energy or surface tension. The content of the paper is

\noindent
$1)$ the definition of the interface free energy from first principles of statistical mechanics; \\
$2)$ a detailed exposition of its  basic properties.

\noindent
We  consider lattice models with short range interactions, like the Ising model. A nice feature of lattice models is that the interface free energy is anisotropic so that some results  are  pertinent to the case of a crystal  in equilibrium with its vapor. The  results of section \ref{section3} hold in full generality.
\end{abstract}

\section{Interface free energy in Statistical mechanical}\label{section2}
\setcounter{equation}{0}

\subsection{Definition of the interface free energy}\label{subsection1}

Consider a physical system {\em at equilibrium} in a vessel $V$, at a first order phase transition point, where two bulk phases, say $A$ and $B$, coexist (for example, an Ising model at zero magnetic field and low temperature). If, when we bring into contact the phases $A$ and $B$, the state of the system is inhomogeneous and  there is spatial separation of the two phases, then  at the common boundary of the two phases emerges a spatially localized structure, called  the interface.
In spite of the low dimensionality of these interfaces and their negligible contribution towards the global overall properties of the physical system their presence is essential for a wealth of important processes in physics, chemistry and biology.
Here we consider only systems at equilibrium, which is a rather severe restriction.

The interface free energy is a thermodynamical quantity and it is best explained when we consider the macroscopic scale, in which the length of the vessel containing the system is the reference length. Under this scale the interface is well-defined and localized. In the case of a flat interface perpendicular to a unit vector ${\bf n}$ it is described mathematically by a plane perpendicular to  ${\bf n}$, which separates the bulk phases, and the state of the system is specified above this plane by giving the value of the order-parameter of one of the bulk phases, say $A$, and below the interface that of the other bulk phase.
The interface free energy or surface tension (per unit area) $\tau({\bf n})$
describes the thermodynamical properties of the interface at equilibrium.
How does one obtain $\tau({\bf n})$ once the interatomic interactions of the system are given? We can answer this question so that we get interesting information about $\tau({\bf n})$ only for few models.
However, the way of defining $\tau({\bf n})$ is quite general and can be  applied in principle to most systems, and its origin can be traced back to the monumental work of J.W. Gibbs,
{\em On the Equilibrium of Heterogeneous Substances} (1875-1878). In statistical mechanics the thermodynamical functions are obtained by computing the partition function. The system is enclosed in a vessel $V$; taking into account the interations of the system with the walls of $V$ we can write an expression for the overall free energy of the system.
The basic postulate  is that we can separate the various contributions to  the overall free energy $F(V)$ at inverse temperature $\beta$ into two parts, up to a small correction term; one part is proportional to the volume of $V$, which is
the bulk free energy of the system, and another one is proportional to the area of the surface of $V$, which is interpreted as the wall free energy. Thus, at a point of first order phase transition, when only phase $A$ is present,
\begin{equation}\label{1.1}
F_A(V)=-\frac{1}{\beta}\ln Z_A(V)=f_{{\rm bulk}}(A)|V| + f_{{\rm wall}}(A)|\partial V|+ o(|\partial V|)\,,
\end{equation}
where $Z_A(V)$ denotes the partition function of the system for phase $A$, $|V|$  the volume of $V$ and  $|\partial V|$ the area of the boundary $\partial V$ of the vessel. A similar expression holds for phase $B$. The bulk terms $f_{{\rm bulk}}(A)$ and $f_{{\rm bulk}}(B)$
are the same because the system is at a first order phase transition point, but  the surface terms $f_{{\rm wall}}(A)$ and $f_{{\rm wall}}(B)$ may be different.
Under specific conditions on the walls, we can obtain (macroscopic) inhomogeneous states with planar interfaces separating the two coexisting bulk phases. In such cases there is an additional contribution to the overall free energy and we {\em postulate} that the free energy can be written as
\begin{equation}\label{1.2}
F_{AB}(V)=-\frac{1}{\beta}\ln Z_{AB}(V)=f_{{\rm bulk}}(AB)|V| + f_{{\rm wall}}(AB)|\partial V|+ \tau({\bf n})|I({\bf n})|+ o(|\partial V|)\,,
\end{equation}
with
$$
f_{{\rm wall}}(AB)=\alpha f_{{\rm wall}}(A)+ (1-\alpha)f_{{\rm wall}}(B)\,.
$$
The term $|I({\bf n})|=O(|\partial V|)$ is the area of the interface perpendicular to the unit vector ${\bf n}$, and $\alpha$ is the proportion of the walls in contact with phase $A$.
Since the system is at a first order transition point $f_{{\rm bulk}}(AB)=f_{{\rm bulk}}(A)=f_{{\rm bulk}}(B)$. Extracting wall free energies is not easy, but it is not necessary to do this if our postulate is correct, because we can eliminate the terms involving $f_{{\rm wall}}(AB)$ and $f_{{\rm wall}}(AB)$ by considering the ratio of partition functions,
\begin{equation}\label{1.3}
-\frac{1}{\beta}\ln\frac{Z_{AB}(V)}{Z_A(V)^{\alpha}Z_B(V)^{1-\alpha}}=\tau({\bf n})|I({\bf n})|+ o(|\partial V|)\,.
\end{equation}
Notice that \eqref{1.3} is always a term of order $O(|\partial V|)$.

An obvious difficulty in getting $\tau({\bf n})$ is that we must know the values of
thermodynamical parameters of the system for which there is phase coexistence. Indeed, for other values of these parameters the system has only one bulk phase and there is no interface. Hence the surface tension is non-zero only for a  specific range of values of the  thermodynamical parameters of the system. This is why in many situations one proceeds differently in Physics. One models directly the interface in order to bypass these problems and then the interface free energy is simply identified with the free energy of the model for which one has standard methods for evaluating it. This is often an adequate way to proceed, but it cannot be applied always. For example when one is studying how the coexisting phases are spatially distributed inside the vessel $V$, we cannot avoid considering
the free energy of interfaces {\em between coexisting phases}.

\subsection{A paradigm, the Ising model}\label{211}

We implement the ideas of section \ref{subsection1} for
the Ising model, for which the mathematical results are the most complete.
We consider the three-dimensional Ising model. The two-dimensional case is also of interest.
Let ${\mathbf Z}^3:=\{t=(t_1,t_2,t_3)\,{:}\; t_i\in{\mathbf Z}\}$ and
$$
\Lambda_{LM}:=\{t\in{\mathbf Z}^3\,{:}\; \max(|t_1|,|t_2|)\leq L\,,\;|t_3|\leq M \}\,.
$$
For each $t\in\Lambda_{LM}$, we introduce a variable $\sigma(t)$, called hereafter spin, which specifies the state of the system at $t$; there are only two possible states labeled by $1$ and $-1$, hence $\sigma(t)=\pm 1$. The configurations of the system are written $\underline{\sigma}$ and are specified by giving for each site $t$ the value of $\sigma(t)$. The spin $\sigma(t)$ interacts with an external real magnetic field $h$, the interaction energy being $-h\,\sigma(t)$, and it interacts with the spin $\sigma(t^\prime)$,
the interaction energy being $-J(t,t^\prime)\,\sigma(t)\sigma(t^\prime)$.
The  energy of the system is equal to
$$
H_{LM}(\underline{\sigma})=-\frac{1}{2}
\sum_{t\in\Lambda_{LM}}\sum_{t^\prime\in\Lambda_{LM}}J(t,t^\prime)\,\sigma(t)\sigma(t^\prime)
-\sum_{t\in\Lambda_{LM}}h\,\sigma(t)\,.
$$
We choose   the coupling constants $J(t,t^\prime)=0$, except if
$t,t^\prime$ are nearest neighbors, in which case $J(t,t^\prime)=J>0$, a fixed value. This interaction favors the alignment of  spins, since the energy is minimal when $\sigma(t)=\sigma(t^\prime)$.
To model the  interaction of the system with the walls, we introduce an inhomogeneous magnetic field $J^\prime\eta(t)$ acting only on the spins located at the boundary of the box $\Lambda_{L,M}$,
$$
W_{LM}^\eta(\underline{\sigma}):=-\sum_{\substack{t\in\Lambda_{LM}:\,|t_3|=M \\
\text{or $\max(|t_1|,|t_2|)=L$}}}J^\prime\eta(t)\sigma(t)\,,
$$
with $J^\prime>0$ and $\eta(t)=1$ or $\eta(t)=-1$, but fixed. Different kind of walls are modeled by  specifying different values for $\eta(t)$ (see below) and choosing different values for $J^\prime$. The overall energy of the system is $H_{LM}+W_{LM}^\eta$.

According to statistical mechanics, the state of the system at equilibrium and at inverse temperature $\beta$ is the Gibbs measure, so that
$$
{\rm Prob}(\underline{\sigma})=\frac{{\rm e}^{-\beta(H_{LM}(\underline{\sigma})+W_{LM}^\eta(\underline{\sigma}))}}{Z_{LM}^\eta}
\quad\text{where}\quad Z_{LM}^\eta=\sum_{\underline{\sigma}^\prime}{\rm e}^{-\beta(H_{LM}(\underline{\sigma}^\prime)+W_{LM}^\eta(\underline{\sigma}^\prime))}\,.
$$
The normalization constant $Z_{LM}^\eta$ is the partition function and
the overall free energy of the system in the box $\Lambda_{LM}$ is
$$
F_{LM}^\eta(\beta,h,J^\prime):=-\frac{1}{\beta}\ln Z_{LM}^\eta\,.
$$

At the thermodynamical limit (infinite-volume limit), the bulk free energy per spin is independent on the choice of $J^\prime>0$ and of $\eta$,
$$
f_{{\rm bulk}}(\beta,h)=\lim_{L\rightarrow\infty}\frac{1}{(2L+1)^d}F_{LL}^\eta(\beta,h,J^\prime)\,.
$$
The model exhibits a first order phase transition with the magnetization as order-parameter
if the external magnetic field $h=0$ and the inverse temperature $\beta>\beta_c(d)$,
where $\beta_c(d)$ is the  inverse critical temperature of the $d$-dimensional Ising model ($0<\beta_c(3)<\beta_c(2)$). For $h=0$ and $\beta>\beta_c(d)$ the spin-flip symmetry of
$H_{LM}$ is broken. There is a phase with positive magnetization $m^*(\beta)$ and another with negative magnetization $-m^*(\beta)$; the bulk free energy  $f_{{\rm bulk}}(\beta,h)$
is not differentiable at $h=0$,
$$
0<m^*(\beta)=\frac{d}{dh}f_{{\rm bulk}}(\beta,h)\Large|_{h=0^+}=-\frac{d}{dh}f_{{\rm bulk}}(\beta,h)\Large|_{h=0^-}\,.
$$
If we choose $\eta(t)\equiv+1$, respectively $\eta(t)\equiv -1$, that is pure boundary conditions, then one can prove that for any value of $J^\prime>0$ the interactions with the walls favor the bulk phase with positive, respectively negative, magnetization.
We are also interested in the case of mixed boundary conditions, which is related to the emergence of a planar interface
perpendicular to ${\bf n}$. Let ${\bf n}=(n_1,n_2,n_3)$. We set
$$
\eta^{\bf n}(t):=\begin{cases} +1 & \text{if $t_1n_1+t_2n_2+t_3n_3\geq 0$,}\\
-1 & \text{if $t_1n_1+t_2n_2+t_3n_3< 0$.}
\end{cases}
$$
Thus $\eta^{{\bf n}}(t)=1$ iff $t$ is above or in the plane $\pi({\bf n})$ perpendicular to ${\bf n}$ and passing through the origin, otherwise  $\eta^{{\bf n}}(t)=-1$. We set $Z_{LM}^{{\bf n}}:=Z_{LM}^{\eta^{\bf n}}$.

\bigskip

\begin{picture}(160,150)(-50,100)
\setlength{\unitlength}{0.7mm}
\put(17,16){\framebox(102,102){}}
\put(20,19){\makebox(0,0){-}}
\multiput(20,25)(0,6){2}{\makebox(0,0){-}}
\multiput(20,37)(0,6){13}{\makebox(0,0){+}}
\multiput(26,19)(6,0){15}{\makebox(0,0){-}}
\put(116,19){\makebox(0,0){-}}
\multiput(116,25)(0,6){13}{\makebox(0,0){-}}
\multiput(116,103)(0,6){3}{\makebox(0,0){+}}
\multiput(20,115)(6,0){16}{\makebox(0,0){+}}
{\thicklines
\put(68,67){\line(3,2){60}}
\put(68,67){\line(-3,-2){60}}
\put(68,67){\vector(-2,3){5}}
}
\put(62,67){${\bf n}$}
\put(124,100){$\pi({\bf n})$}
\put(72,5){\makebox(0,0){The field $J^\prime\eta^{\bf n}$ acting on the boundary spins of $\Lambda_{L,L}$.}}
\end{picture}

\vspace*{3.5truecm}

\subsubsection{Macroscopic limit and definition of $\tau$ in the Ising model}\label{212}

We suppose that the external magnetic field $h=0$ and $\beta>\beta_c(d)$, $d=2,3$. We choose $0<a<1$ and we partition the set $\Lambda_{LL}$ into cubic cells $C_i$ of linear size $L^a$; for each cell $C_i$ we introduce the empirical
magnetization  (averaged magnetization over the cell $C_i$)
$$
m_{C_i}(\underline{\sigma}):=|C_i|^{-1}\sum_{t\in C_i}\sigma(t)\,.
$$
We scale all lengths by $L^{-1}$,
so that the distance between neighboring spins is $L^{-1}$. In the fixed reference macroscopic box
$V=\{(x_1,x_2,x_3)\in{\mathbf R}^d\,{:}\; |x_i|\leq 1\}$ we make a coarse-grained description of  the state of the system by specifying
the coarse-grained magnetization profile,
$$
\rho_L(x|\underline{\sigma}):=m_C(\underline{\sigma})\quad\text{if $(Lx_1,Lx_2,Lx_3)\in C_i$, $\forall$
$(x_1,x_2,x_3)\in V$.}
$$
By definition
the probability of the profile $\rho_L(x|\underline{\sigma})$ is  the joint probability of the block-spins $m_{C_i}(\underline{\sigma})$ induced by the
Gibbs measure.
The macroscopic limit is the limit when $L^{-1}$ tends to $0$.  If we choose $\eta\equiv +1$, when $L^{-1}$ tends to $0$, then for {\em any value of} $J^\prime>0$ the probability measure
on the density profiles
becomes concentrated
on the unique magnetization profile $\rho(x)\equiv m^*(\beta)$; this constant profile describes the macroscopic state of the $+$-phase of the model. The macroscopic limit corresponds to the regime of the law of large numbers in probability theory.

To obtain $\tau({\bf n})$ we use  \eqref{1.3} and the fact that by symmetry of the model $Z_{LL}^+=Z_{LL}^-$, so that we only need to compare
$Z_{LL}^+$  and $Z_{LL}^{{\bf n}}$. We set
\begin{equation}\label{tau}
\tau({\bf n}):=-\frac{1}{\beta|\pi({\bf n})\cap V|}\lim_{L\rightarrow\infty}\frac{1}{L^{d-1}}\ln\frac{Z_{LL}^{{\bf n}}}{Z_{LL}^+}\,.
\end{equation}
The term $|\pi({\bf n})\cap V|$ is the area of the intersection of the plane $\pi({\bf n})$ with $V$.
One can prove: {\em the limit \eqref{tau} is independent on $J^\prime\geq J$, and for $\beta>\beta_c(d)$ the function $\tau({\bf n})$ verifies properties a), b) and c) of \eqref{abc}; in the macroscopic limit the measure on the density profiles is concentrated on
the unique magnetization profile}
$$
\rho_{{\bf n}}(x):=\begin{cases}+ m^*(\beta)& \text{if $x$ is above $\pi({\bf n})$}\\
-m^*(\beta)& \text{if $x$ is below $\pi({\bf n})$.}
\end{cases}
$$
This profile describes a macroscopic state with a planar interface perpendicular to ${\bf n}$. Therefore $\tau({\bf n})$ can be interpreted as the free energy of that interface perpendicular to ${\mathbf n}$.

The condition $J^\prime\geq J$  is important, because for some values of $J^\prime<J$ and $\beta$ the physics near the walls of the system is different: a surface phase transition may take place and
portions of the interface  may be pinned to the walls. As a consequence of this phenomenon
in the macroscopic limit the interaction of the system with the walls  given by $\eta^{\bf n}$  {\em may not induce an interface  perpendicular to ${\bf n}$}.  For example, in the two-dimensional case, the macroscopic state may have an interface making an angle with the vertical walls of the vessel, whose value is given by the Young-Herring equation, so that (\ref{tau}) may  not be equal to  $\tau({\bf n})$, or, if $J^\prime$ is small enough and the macroscopic box is a square, then the whole interface may even be pinned to the walls, so that there is no interface through the macroscopic box.
In  such cases the limit (\ref{tau}) depends on $J^\prime$.
The condition $J^\prime\geq J$ has a simple physical interpretation; it ensures that the walls of the box $V$ are in the complete wetting regime, so that the interface cannot be pinned to the walls.
In the literature the standard choice for ferromagnetic models is $J^\prime=J$, so that (\ref{tau}) gives the correct definition of $\tau({\bf n})$. These results illustrate the fact that it is important to choose correctly the interactions of the system with the walls in order to use definition (\ref{1.3}). One must avoid the possibility of pinning of the interface to the walls. Any wall interactions which induce a macroscopic state with an interface perpendicular to ${\bf n}$ and such that otherwise (\ref{1.3}) is independent of the chosen interactions are admissible for defining
the interface free energy.

Several other definitions for $\tau({\bf n})$ have been proposed for the Ising or similar models. Most of them  involve a ratio  of partition functions and are based on the same pattern leading to (\ref{tau}) so that we shall not review them here (see section \ref{section4} for references). A possibility of avoiding the above problem with the walls is to suppress (partially) the walls of the system by taking (partial) periodic boundary conditions. Then one imposes a condition implying the existence of a single planar interface perpendicular to ${\bf n}$. There are also variants of (\ref{tau}) where one considers a box $\Lambda_{LM}$ instead of $\Lambda_{LL}$ and take first the limit $M\rightarrow\infty$ before taking $L\rightarrow\infty$. When $J^\prime<J$ this limit may give a different answer as the limit \eqref{tau}. On the other hand,  if $J^\prime\geq J$, then one can take the limits in any order, first $L\rightarrow\infty$ and then $M\rightarrow\infty$ or vice-versa, or simultaneously $L\rightarrow\infty$ and $M\rightarrow\infty$. The reason is that the walls are in the complete wetting regime and the interface is not pinned to the walls.
In general it is not easy to show that reasonable definitions give the same value for $\tau({\bf n})$.

The surface tension for the two-dimensional Ising model can be computed exactly. Onsager computed the interface free energy for ${\bf n}=(0,1)$,
$$
\beta\tau((0,1))=2(K-K^*)\,,\;\beta>\beta_c(2)\quad\text{and}\quad \tau((0,1))=0\,,\;\text{otherwise,}
$$
where $K^*$ is defined by $\exp(-2K^*)=\tanh K$ and $K=\beta J$.
Onsager did not use the definition \eqref{tau}; the computation of $\tau((0,1))$ defined by \eqref{tau} is due to Abraham and Martin-L\"of. The full interface free energy has been computed by McCoy and Wu.

\subsubsection{Inequalities for  $\tau$ in the Ising model}\label{213}

We set in this subsection
$$
{\bf n}(\theta):=(0,-\sin\theta,\cos\theta)\quad\text{and}\quad \tau(\theta):=\tau({\bf n}(\theta))\,.
$$
There are two inequalities which relate the interface free energy and the order-parameter,
\begin{equation}\label{above}
\tau(0)\leq 2(m^*(\beta))^2\,,
\end{equation}
and
\begin{equation}\label{below}
\frac{d(\beta\tau(0))}{d\beta}\geq 2(m^*(\beta))^2\,.
\end{equation}
These inequalities indicate that $\tau(0)>0$   iff there is a phase transition.  Since $\tau(0)$ is a minimal value, we have $\tau({\bf n})>0$ iff $h=0$ and $\beta>\beta_c(d)$.

We now introduce the {\bf step free energy} $\tau_{{\rm step}}$. This quantity is interesting only for $d=3$. Let $\eta^*(t)$ be defined as
$$
\eta^*(t):=\begin{cases}+1 &\text{if $t_3>0$ or if $t_3=0$ and $t_2\geq 0$,}\\
-1 & \text{if $t_3<0$ or if $t_3=0$ and $t_2<0$.}
\end{cases}
$$
By definition, if $d=3$,
$$
\tau_{{\rm step}}:=-\frac{1}{\beta}\lim_{L\rightarrow\infty}\frac{1}{2L+1}\ln\frac{Z_{LL}^{\eta^*}}{Z_{LL}^{{\bf n}^*}}\quad\text{with ${\bf n}^*=(0,0,1)$}\,.
$$
One has $\tau_{{\rm step}}\geq 0$ and we state  two important inequalities. The first inequality
\begin{equation}\label{step}
\tau(\theta)-\tau(0)\geq |\sin\theta| \tau_{{\rm step}}
\end{equation}
gives information about the  non-differentiability of $\tau$ at $\theta=0$. Positivity of $\tau_{{\rm step}}$ implies that $\tau(\theta)$ is not differentiable at $\theta=0$ since $\tau$ has a minimum at $\theta=0$ and
$$
\lim_{\theta\downarrow 0}\frac{d}{d\theta}\tau(\theta)\geq \tau_{{\rm step}}\,.
$$
The physical consequence of the  non-differentiability of $\tau$ is explained  at the end of section \ref{233}; non-differentiability of $\tau$   implies the existence of a facet for the equilibrium shape $W_\tau$ defined in \eqref{wulff}.
The second inequality relates the step free energy of the three-dimensional model to the two-dimensional interface free energy, which we write here $\tau_2(\theta)$,
\begin{equation}\label{comparison}
\tau_{{\rm step}}\geq \tau_2(0)\,.
\end{equation}
Therefore, if $\beta>\beta_c(2)$ then $\tau_{{\rm step}}>0$.

\section{Basic properties of the interface free energy}\label{section3}
\setcounter{equation}{0}

We examine in this section what are the basic properties of $\tau$ and we discuss the thermodynamical stability of interfaces.

\subsection{Convexity of the interface free energy}\label{subsection2.2}

We assume that ${\tau}({\bf n})>0$ for each unit vector ${\bf n}$.
We consider the three-dimensional case. Elements of the Euclidean space ${\mathbf E}_3$ are written
${\bf x}=(x_1,x_2, x_3)$; we denote the Euclidean scalar product by
$$
\langle\,{\bf x}|{\bf y}\,\rangle:=x_1y_1+x_2y_2+x_3y_3\,,
$$
and the Euclidean norm by  $\|{\bf x}\|$.

By convention $\tau({\bf n})$, with $\|{\bf n}\|=1$, is the physical value of the interface free energy of an interface perpendicular to ${\bf n}$. It is convenient to extend the definition of $\tau$ to ${\mathbf E}_3$, as a positively homogeneous function,  by setting
$$
\tau({\bf x}):=\|{\bf x}\|\tau({\bf x}/\|{\bf x}\|)\,.
$$
We introduce the half-space
$$
H({\bf n}):=\{{\bf x}\,{:}\; \langle\, {\bf x}|{\bf n}\,\rangle\leq \tau({\bf n})\}\,,
$$
whose boundary is the plane $\partial H({\bf n})=\{{\bf x}\,{:}\; \langle\, {\bf x}|{\bf n}\,\rangle= \tau({\bf n})\}$. Notice that
$H({\bf n})=H({t\bf n})$ for all $t>0$. Besides we define the {\bf equilibrium shape} $W_\tau$ as the
convex set, which is the intersection of the half-spaces $H({\bf n})$, that is
\begin{equation}\label{wulff}
W_\tau:=\{{\bf x}\,{:}\; \langle\, {\bf x}|{\bf n}\,\rangle\leq \tau({\bf n})\,,\;\forall\, {\bf n}\}\,.
\end{equation}

The next argument  is due to Herring \cite{H1}; it is reproduced almost in its original form.
Let  ${\mathcal T}(A_0,A_1,A_2,A_3)$ be the tetrahedron with vertices $A_0,A_1,A_2,A_3$. The face opposite to the vertex $A_i$ is denoted by $\Delta_i$; it contains all vertices $A_j$, $j\not=i$, and its area is $|\Delta_i|$. Let ${\bf n}_i$ be the outward unit normal to the face $\Delta_i$. We have
$$
|\Delta_0|{\bf n}_0+|\Delta_1|{\bf n}_1+|\Delta_2|{\bf n}_2+|\Delta_3|{\bf n}_3=0\,.
$$
Set
$$
{\bf n}:=-{\bf n}_0=\frac{|\Delta_1|}{|\Delta_0|}{\bf n}_1+\frac{|\Delta_2|}{|\Delta_0|}{\bf n}_2+\frac{|\Delta_3|}{|\Delta_0|}{\bf n}_3\,.
$$
We compare the free energies $|\Delta_0|\tau({\bf n})$ and $|\Delta_1|\tau({\bf n}_1)+|\Delta_2|\tau({\bf n}_2)+|\Delta_3|\tau({\bf n}_3)$, or which is the same, $\tau({\bf n})$ and
$$
\frac{|\Delta_1|}{|\Delta_0|}\tau({\bf n}_1)+\frac{|\Delta_2|}{|\Delta_0|}\tau({\bf n}_2)+\frac{|\Delta_3|}{|\Delta_0|}\tau({\bf n}_3)\,.
$$
Let ${\bf m}_1$, ${\bf m}_2$, ${\bf m}_3$ be reciprocal vectors to ${\bf n}_1$, ${\bf n}_2$, ${\bf n}_3$, defined by the relations $\langle\,{\bf m}_i|{\bf n}_j\,\rangle=0$ if $i\not=j$ and
$\langle\,{\bf m}_i|{\bf n}_i\,\rangle=1$ otherwise. We have $\langle\,{\bf m}_i|{\bf n}\rangle=|\Delta_i|/|\Delta_0|$, so that
\begin{equation*}
\sum_{i=1}^3\frac{|\Delta_i|}{|\Delta_0|}\tau({\bf n}_i)=
\langle\,\sum_{i=1}^3\tau({\bf n}_i){\bf m}_i|{\bf n}\,\rangle\equiv\langle\,{\bf y}|{\bf n}\,\rangle\,.
\end{equation*}
The vector ${\bf y}=\sum_{i=1}^3\tau({\bf n}_i){{\bf m}_i}$ verifies the identities
$\langle\,{\bf y}|{\bf n}_i\,\rangle=\tau({\bf n}_i)$, $i=1,2,3$, i.e. ${\bf y}$ is the intersection point of the three planes $\partial H({\bf n}_1)$, $\partial H({\bf n}_2)$ and $\partial H({\bf n}_3)$. We have three cases.

\noindent
$(1)$. If $\langle\,{\bf y}|{\bf n}\,\rangle<\tau({\bf n})$, then the intersection of the plane $\partial H({\bf n})$
with $W_\tau$ is empty, since $W_\tau$ is a subset of $H({\bf n}_1)\cap H({\bf n}_2)\cap H({\bf n}_3)$.
A (hypothetical) planar interface perpendicular to ${\bf n}$ with interface free energy $\tau({\bf n})$  can be deformed using the tetrahedron ${\mathcal T}(A_0,A_1,A_2,A_3)$ by lowering its free energy.
Hence, at nonzero temperature such a planar interface with interface free energy $\tau({\bf n})$ is  unstable
and  cannot exist {\em at equilibrium} since the free energy is minimal. Therefore we must have
\begin{equation}\label{2.2}
|\Delta_0|\tau({\bf n})\leq |\Delta_1|\tau({\bf n}_1)+|\Delta_2|\tau({\bf n}_2)+|\Delta_3|\tau({\bf n}_3)\,.
\end{equation}

\noindent
$(2)$. The case $\langle\,{\bf y}|{\bf n}\,\rangle=\tau({\bf n})$ is a borderline case. Here the intersection of $\partial H({\bf n})$ with $W_\tau$ is at most one point. If $\partial H({\bf n})\cap W_\tau=\emptyset$, then one also show that the interface is also unstable and cannot exist at equilibrium.
If
$\partial H({\bf n})\cap W_\tau=\{{\bf y}\}$, then ${\bf y}\in W_\tau\cap\partial H({\bf n}_1)\cap\partial H({\bf n}_2)\cap\partial H({\bf n}_3)$ is called a {\bf corner} of $W_\tau$. In this case $\partial H({\bf n})$ is a support plane of $W_\tau$ at ${\bf y}$, that is $W_\tau\subset H({\bf n})$ and $W_\tau\cap \partial H({\bf n})\ni {\bf y}$, but it is not a tangent plane of $W_\tau$. Notice that if $W_\tau$ has no corner, then at equilibrium we only have the next case $(3)$.

\noindent
$(3)$. If $\langle\,{\bf y}|{\bf n}\,\rangle>\tau({\bf n})$, then
\begin{equation}\label{2.3}
|\Delta_0|\tau({\bf n})<|\Delta_1|\tau({\bf n}_1)+|\Delta_2|\tau({\bf n}_2)+|\Delta_3|\tau({\bf n}_3)\,.
\end{equation}

We can repeat the same argument with a right prism whose base is a triangle with vertices $A_0,A_1,A_2$ and whose length is very large. Let $\ell_i$ be the side of the triangle opposite to the vertex $A_i$, $|\ell_i|$ be its length, and ${\bf n}_i$ the outward unit normal to the side $\ell_i$ in the plane of the triangle. Again we have
$$
|\ell_0|{\bf n}_0+|\ell_1|{\bf n}_1+|\ell_2|{\bf n}_2=0\,,
$$
and we set ${\bf n}:=-{\bf n}_0=|\ell_1|/|\ell_0|{\bf n}_1+ |\ell_2|/|\ell_0|{\bf n}_2$.
In the plane spanned by ${\bf n}_1$ and ${\bf n}_2$ let ${\bf m_1}$ and ${\bf m_2}$ be reciprocal vectors to ${\bf n}_1$ and ${\bf n}_2$. Then
$$
\sum_{i=1}^2\frac{|\ell_i|}{|\ell_0|}\tau({\bf n}_i)=
\langle\,\sum_{i=1}^2\tau({\bf n}_i){\bf m}_i|{\bf n}\,\rangle\equiv\langle\,{\bf z}|{\bf n}\rangle\,.
$$
The vector ${\bf z}=\sum_{i=1}^2\tau({\bf n}_i){\bf m}_i$
belongs to the intersection of the planes
$\partial H({\bf n}_1)$ and $\partial H({\bf n}_2)$. We can draw similar conclusions as above.
If $\langle\,{\bf z}|{\bf n}\rangle<\tau({\bf n})$, then the interface $\partial H({\bf n})$ with interface free energy $\tau({\bf n})$ is unstable and does not exist at equilibrium. Hence, at equilibrium,
\begin{equation}\label{2.5}
|\ell_0|\tau({\bf n})\leq|\ell_1|\tau({\bf n}_1)+|\ell_2|\tau({\bf n}_2)\,.
\end{equation}
If $\langle\,{\bf z}|{\bf n}\rangle=\tau({\bf n})$, then either $\partial H({\bf n})\cap W_\tau=\emptyset$ or  $\partial H({\bf n})\cap W_\tau\not =\emptyset$; in the latter case $W_\tau$ has an edge or a corner and $\partial H({\bf n})$ is a support plane (but not a tangent plane of $W_\tau$). If $\langle\,{\bf z}|{\bf n}\rangle>\tau({\bf n})$, then
\begin{equation}\label{2.4}
|\ell_0|\tau({\bf n})<|\ell_1|\tau({\bf n}_1)+|\ell_2|\tau({\bf n}_2)\,.
\end{equation}

Since $\tau$ has been defined as a positively homogeneous function, it is immediate to see that \eqref{2.5}  for all choices of ${\bf n}_1$, ${\bf n}_2$, $\ell_1$ and $\ell_2$  is equivalent to
\begin{equation}\label{2.6}
\tau({\bf x}+{\bf y})\leq\tau({\bf x})+\tau({\bf y})\quad \forall \,{\bf x},{\bf y}\,.
\end{equation}
From \eqref{2.6} one easily gets \eqref{2.2} for any tetrahedron. Conversely, if \eqref{2.2} is valid for any tetrahedron, then we get \eqref{2.6} by deforming a tetrahedron into a prism, sending one of the vertex to infinity. The important inequality is \eqref{2.6} (or \eqref{2.5}).

{\em To summarize, for two distinct phases at equilibrium, the interface free energy
is a continuous convex function, which is positive and sublinear, that is
\begin{eqnarray}
\text{{\rm (a)}}&\quad &\tau({\bf x})>0\quad {\bf x}\not=0\,,\nonumber\\
\text{{\rm (b)}}&\quad &\tau(t{\bf x})=t\, \tau({\bf x})\quad\text{$\forall {\bf x}$ and all $t\geq 0$}\,,  \label{abc} \\
\text{{\rm (c)}}&\quad &\tau({\bf x}+{\bf y})\leq\tau({\bf x})+\tau({\bf y})\quad \text{$\forall \,{\bf x},{\bf y}$.}\nonumber
\end{eqnarray}
By a classical result of Minkowski it is the support function of the convex set $W_\tau$, that is
\begin{equation}\label{support}
\tau({\bf x})=\sup\{\langle\,{\bf x}|{\bf y}\,\rangle\,{:}\; {\bf y}\in W_\tau\}\,.
\end{equation}
}

\noindent
We say that an {\bf interface perpendicular to ${\bf n}$ is  (thermodynamically) stable} if
\begin{equation}\label{stable}
\tau({\bf x}+{\bf y})<\tau({\bf x})+\tau({\bf y})\quad \text{$\forall \,{\bf x},{\bf y}$ linearly independent, such that
${\bf x}+{\bf y}={\bf n}$}\,.
\end{equation}
In that case  \eqref{2.3} and \eqref{2.4} hold.
In general the choice of the normal to the interface does not matter, so that we also have $\tau({\bf n})=\tau(-{\bf n})$. This means that $\tau({\bf n})$ defines a norm on ${\mathbf E}_3$.

\subsection{Interface free energy  and equilibrium shape $W_\tau$}\label{subsection2.3}

We suppose that $\tau$ is given, verifying properties a), b) and c) of \eqref{abc}
(we do not assume that $\tau({\bf n})=\tau(-{\bf n})$).
Under these assumptions $W_\tau$ is a {\bf convex body}, i.e.  $W_\tau\subset{\mathbf E}_3$ is a
bounded, closed convex set with non-empty interior. The point $0$ is an interior point of $W_\tau$, because the continuity of $\tau$ implies that $\tau({\bf n})\geq a>0$ for all ${\bf n}$, $\|{\bf n}\|=1$.

\subsubsection{Equilibrium shape $W_\tau$ and its polar dual $W^*_\tau$}\label{231}

A {\bf face} $F$ of a convex set $K\subset {\mathbf E}_3$ is a convex subset $F\subset K$,  such that
$$
{\bf x}\,,\,{\bf y}\in K\;\text{and}\;\frac{1}{2}({\bf x}+{\bf y})\in F\quad\text{imply}\quad
{\bf x}\,,\,{\bf y}\in F\,.
$$
Two-dimensional faces are called {\bf facets} of $K$ and one-dimensional faces are called {\bf edges} of $K$. An {\bf extremal point ${\bf z}$} of $K$ is a face $F=\{{\bf z}\}$. Equivalently, ${\bf z}$
is an extremal point if it cannot be written as ${\bf z}=\lambda {\bf x}+(1-\lambda){\bf y}$ with
${\bf x},{\bf y}\in K$ and $\lambda\in (0,1)$, except by taking ${\bf z}={\bf x}={\bf y}$.
The set of extremal points is denoted by ${\rm ext}K$. Minkowski's theorem states that {\em a convex body $K$ is equal to the convex hull of its extremal points, that is}
$$
K=\left\{{\bf x}\,{:}\; \sum_{i=1}^k\lambda_i{\bf x}_i\,{:}\; \lambda_i\geq 0\,,\;\sum_{i=1}^k\lambda_i=1\,,\;
{\bf x}_i\in {\rm ext}K\;\forall\,i\,,\;k \;\text{arbitrary}\right\}\,.
$$

\noindent
{\bf Remark.\,} In dimension two  we call ''facets`` the one-dimensional faces of $K\subset {\mathbf E}_2$.

To study the convex set  $W_\tau$ we introduce its polar dual $W^*_\tau$.
The definition of $W^*_\tau$ is based on the dual relationship between non-zero vectors ${\bf v}$ and
closed half-spaces ${\bf v}^*$ containing the origin,
$$
{\bf v}^*:=\{{\bf x}\,{:}\; \langle\,{\bf v}|{\bf x}\,\rangle\leq 1\}\,.
$$
The {\bf polar dual}  or {\bf polar set} $W^*_\tau$ of $W_\tau$ is
$$
W_\tau^*:=\bigcap\{{\bf x}^*\,{:}\; {\bf x}\in W_\tau\}=
\{{\bf u}\,{:}\; \langle\,{\bf x}|{\bf u}\,\rangle\leq 1\quad\forall\,{\bf x}\in W_\tau\}\,.
$$
From the definition it is immediate that $W_\tau^*$ is a convex set.
It is a closed set, since the scalar product is continuous, and $0$ is an interior point of $W_\tau^*$.
Hence $W^*_\tau$ is a convex body. Since for ${\bf x}\in W_\tau$ we have $\langle\,{\bf x}|{\bf u}\,\rangle\leq 1$ for all ${\bf u}\in W^*_\tau$, we deduce that $W_\tau^{**}\supset W_\tau$.
Since $W_\tau$ is a convex body having the origin as an interior point, we also have $W_\tau=W_\tau^{**}$. Indeed,
suppose that ${\bf z}\not\in W_\tau$. Since $W_\tau$ is a closed convex set, there exists a plane $\{{\bf x}\,{:}\; \langle\,{\bf x}|{\bf u}\,\rangle=\alpha\}$ which separates ${\bf z}$ and $W_\tau$, i.e.
$$
W_\tau\subset \{{\bf x}\,{:}\; \langle\,{\bf x}|{\bf u}\,\rangle\leq \alpha\}\quad\text{and}\quad
\langle\,{\bf z}|{\bf u}\,\rangle>\alpha\,.
$$
Since $0$ is an interior point of $W_\tau$, we can choose ${\bf u}$ so that $\alpha>0$. Therefore,
for all ${\bf y}\in W_\tau$, $\langle{\bf y}|{\bf u}/\alpha\,\rangle\leq 1$, i.e. ${\bf u}/\alpha\in W^*_\tau$. Since $\langle{\bf z}|{\bf u}/\alpha\,\rangle>1$ we conclude that ${\bf z}\not\in W^{**}_\tau$, so that
$$
W_\tau=W_\tau^{**}\,.
$$
There is a simple description of $W^*_\tau$ in terms of $\tau$. If $\tau({\bf u})\leq 1$, it follows from  \eqref{wulff} that ${\bf u}\in W^*_\tau$.
Conversely, since $\tau$ is the support function of $W_\tau$,  there exists
for any ${\bf u}$ a point
${\bf z}\in W_\tau$ such that $\langle\,{\bf z}|{\bf u}\,\rangle=\tau({\bf u})$. Therefore, when ${\bf u}\in W^*_\tau$, we have $\tau({\bf u})=\langle\,{\bf z}|{\bf u}\,\rangle\leq 1$.
Hence
\begin{equation}\label{polar}
W^*_\tau=\{{\bf u}\,{:}\; \tau({\bf u})\leq 1\}\quad\text{and}\quad \tau({\bf x})=\min\{t\geq 0\,{:}\; {\bf x}/t\in W^*_\tau\}\,.
\end{equation}
The statements in \eqref{polar} mean that $\tau$ is the gauge function of $W^*_\tau$. We can interpret the interface free energy  either as the support function of $W_\tau$, or as the gauge function of $W^*_\tau$. The boundary $\partial W^*_\tau$ of the polar set is simply the level-$1$ surface of  $\tau$. Since $\tau$ is positively homogeneous, the level-$t$ surface is obtained from the level-$1$ surface by a dilation of factor $t$. Since $(\partial W^*_\tau)^*=
W_\tau^{**}$ and ${\bf n}^*=H({\bf n})$ for any ${\bf n}\in \partial W^*_\tau$, the boundary points of $W^*_\tau$ give a natural labeling of the support planes of $W_\tau$ (see also section \ref{232}).

{
\includegraphics[width=6cm]{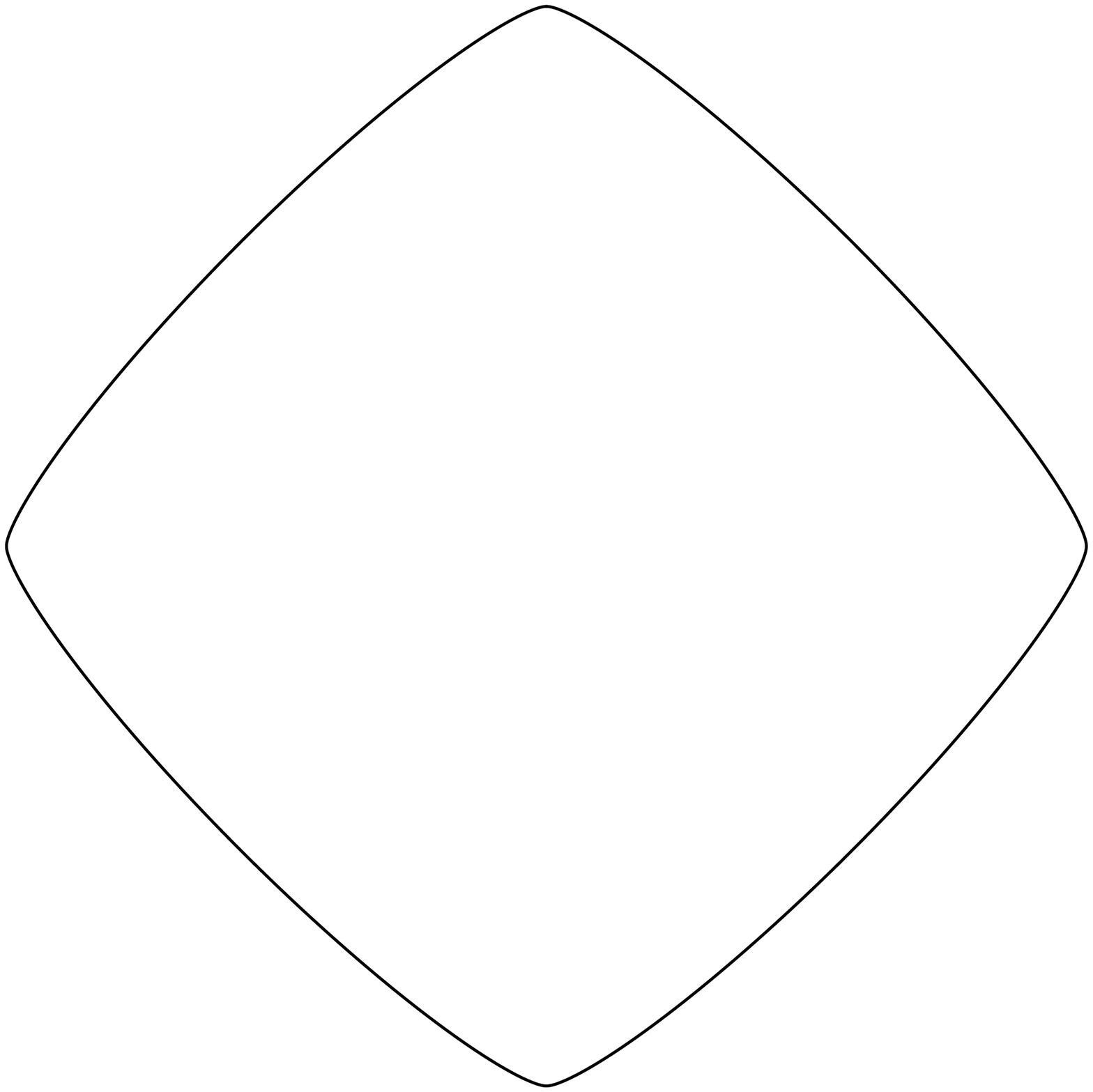}
\hspace{2cm}
\includegraphics[width=6cm]{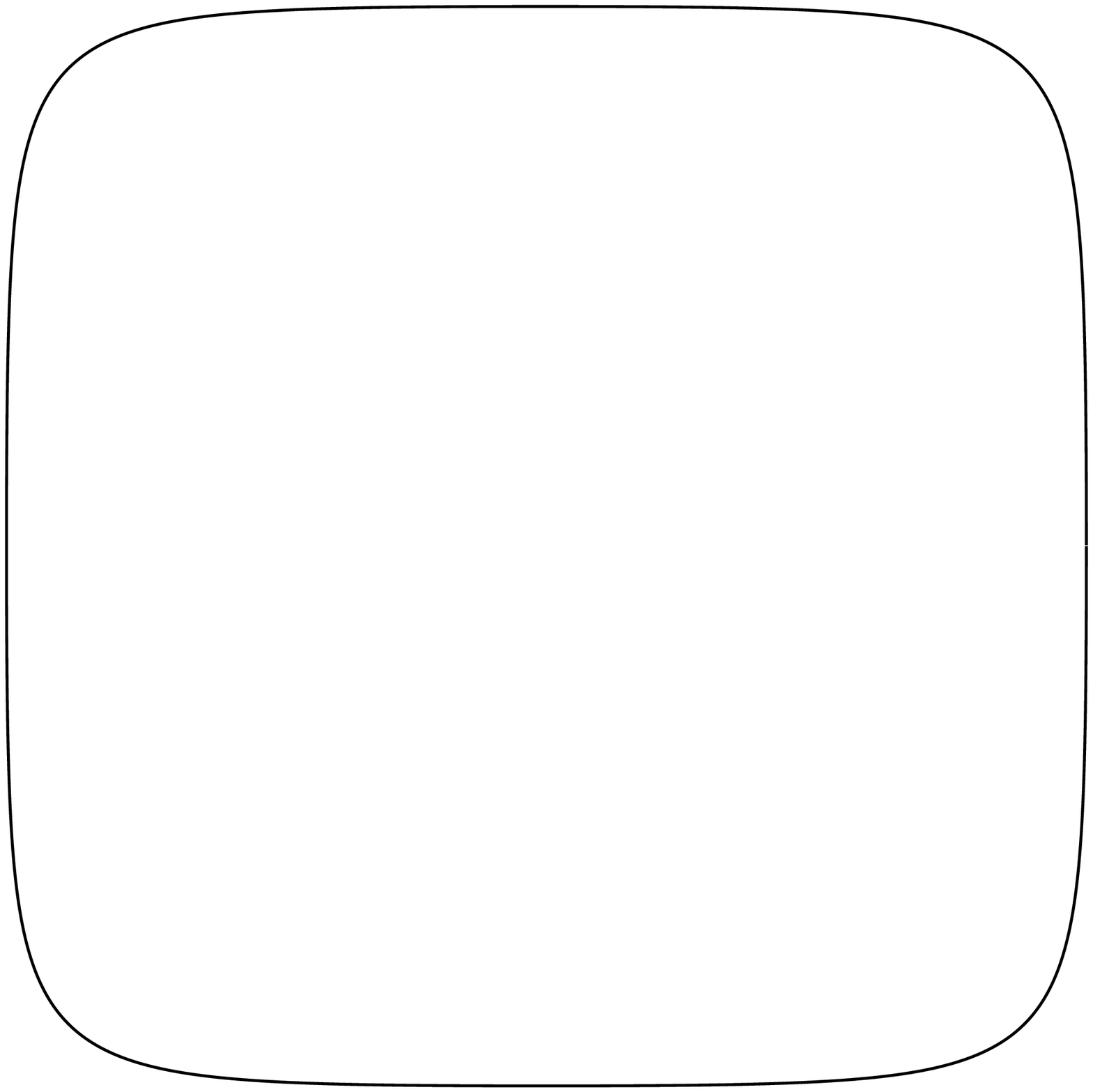}

\centerline{Polar set and equilibrium shape. 2D Ising model $J=1$, $\beta=3$.}
}

\noindent
{\bf Remark.\,} Since $W_\tau=W_\tau^{**}$ we can exchange the roles of $W_\tau$ and $W^*_\tau$.
Let $\rho$ be the support function of $W^*_\tau$. This function is the gauge function of
the equilibrium shape $W_\tau$, so that
$$
W_\tau=\{{\bf x}\,{:}\; \rho({\bf x})\leq 1\}\,.
$$
By definition of the support function (see \eqref{support}) we have
$$
\rho({\bf x})=
\sup\{\langle\,{\bf x}|{\bf y}\,\rangle\,{:}\; {\bf y}\in W^*_\tau\}=\sup_{{\bf y}\not=0}\frac{\langle\,{\bf x}|{\bf y}\,\rangle}{\tau({\bf y})}\quad\implies\quad \langle\,{\bf x}|{\bf y}\,\rangle\leq \rho({\bf x})\tau({\bf y})\,.
$$

\noindent
{\bf Example.\,} The H\"older norms on ${\mathbf E}_3$ are defined by
$$
\|{\bf x}\|_p:=\begin{cases}\left(\sum_{i=1}^3|{\bf x}_i|\right)^{1/p}&\text{if $1\leq p<\infty$,}\\
\max\{|{\bf x}_i|\,,\,i=1,2,3\}& \text{if $p=\infty$.}
\end{cases}
$$
Suppose that $\tau({\bf x})=\|{\bf x}\|_p$. Then
$$
W_\tau=\{{\bf x}\,{:}\; \|{\bf x}\|_q\leq 1\}\quad\text{and}\quad W^*_\tau=\{{\bf x}\,{:}\; \|{\bf x}\|_p\leq 1\}\,,
$$
where $q$ verifies $1/p+1/q=1$.

There is another way of expressing the equilibrium shape using $\tau$.
For any ${\bf y}$ we introduce the affine functional
$$
{\bf x}\mapsto \phi_{\bf y}({\bf x}):=\langle\,{\bf x}|{\bf y}\,\rangle-\tau({\bf y})\,.
$$
The half-space $H({\bf n})=\{{\bf x}\in{\mathbf R}^3\,{:}\; \phi_{\bf n}({\bf x})\leq 0\}$.  The equilibrium shape can be written
$$
W_\tau=\{{\bf x}\,{:}\; \sup_{\bf y}\phi_{\bf y}({\bf x})\leq 0\}\,.
$$
Since $\tau$ is sublinear it is easy to verify that
$$
\tau^*({\bf x}):=\sup_{\bf y}(\langle\,{\bf x}|{\bf y}\,\rangle-\tau({\bf y}))=
\begin{cases} 0 & \text{if ${\bf x}\in W_\tau$}\\
\infty & \text{otherwise.}
\end{cases}
$$
Thus {\em the Legendre transform $\tau^*$ of the interface free energy is the indicator function of the equilibrium shape $W_\tau$}.

\subsubsection{Extremal support planes and stability of interfaces}\label{232}

We say that  $\partial H({\bf n})$ is an {\bf extremal support plane} of $W_\tau$ iff $\phi_{{\bf n}}({\bf x})$ cannot be written as
$$
\phi_{{\bf n}}({\bf x})=c_1\phi_{{\bf n}_1}({\bf x})+c_2\phi_{{\bf n}_2}({\bf x})\quad c_1>0\,,\;c_2>0\,,
$$
except by taking ${\bf n_1}=t_1{\bf n}$ and ${\bf n}_2=t_2{\bf n}$, $t_1>0$ and $t_2>0$.
Let $H({\bf n}_1)\not=H({\bf n}_2)$, $c_1>0$ and $c_2>0$ be given.
Let ${\bf n}=c_1{\bf n}_1+c_2{\bf n}_2$ and ${\bf x}\in W_\tau$. Then, by sublinearity of $\tau$,
$$
0\geq \phi_{{\bf n}}({\bf x})=\langle\,{\bf x}|{c_1{\bf n}_1+c_2{\bf n}_2}\,\rangle-\tau(c_1{\bf n}_1+c_2{\bf n}_2)
 \geq c_1\phi_{{\bf n}_1}({\bf x})+c_2\phi_{{\bf n}_2}({\bf x})\,.
$$
From this
we conclude that $\partial H({\bf n})$ is extremal iff
$$
\tau({\bf n})<c_1 \tau({\bf n}_1)+c_2 \tau({\bf n}_2)\quad\text{$\forall\,{\bf n}_1,{\bf n}_2$ linearly independent, such that $c_1{\bf n}_1+c_2{\bf n}_2={\bf n}$}\,,
$$
that is, iff the interface perpendicular to ${\bf n}$ is stable (see \eqref{stable}). When the support planes $\partial H({\bf n})$ are parametrized by ${\bf n}\in \partial W^*_\tau$, $\partial H({\bf n})$ is extremal iff ${\bf n}$ is an extremal point of $W^*_\tau$. Indeed,
if ${\bf n}=\lambda{\bf n}_1+(1-\lambda){\bf n}_2$ is a non-extremal boundary point of $W^*_\tau$, then
$$
1=\tau({\bf n})\leq \lambda\tau({\bf n}_1)+(1-\lambda)\tau({\bf n}_2)\leq 1\;\implies\;
\tau({\bf n})=\lambda\tau({\bf n}_1)+(1-\lambda)\tau({\bf n}_2)\,,
$$
so that $\partial H({\bf n})$ is non-extremal. Conversely, if  $\partial H({\bf n})$ is non-extremal, then there exist $c_1>0$, $c_2>0$, ${\bf n}_1$ and ${\bf n}_2$ so that ${\bf n}=c_1{\bf n}_1+c_2{\bf n}_2$ and $\tau({\bf n})=c_1 \tau({\bf n}_1)+c_2 \tau({\bf n}_2)$. Putting ${\bf u}={\bf n}/\tau({\bf n})$, ${\bf u}_1={\bf n}_1/\tau({\bf n}_1)$ and ${\bf u}_2={\bf n}_2/\tau({\bf n}_2)$, we get
$$
{\bf u}=
\frac{c_1\tau({\bf n}_1)}{\tau({\bf n})}{\bf u}_1+\frac{c_2\tau({\bf n}_2)}{\tau({\bf n})}{\bf u}_2\quad
\text{and}\quad\frac{c_1\tau({\bf n}_1)}{\tau({\bf n})}+\frac{c_2\tau({\bf n}_2)}{\tau({\bf n})}=1\,,
$$
so that ${\bf u}\in\partial W^*_\tau$ is non-extremal. To summarize, {\em the support plane $\partial H({\bf n})$ of $W_\tau$, with ${\bf n}\in \partial W^*_\tau$, is extremal iff ${\bf n}$ is an extremal point  of $W^*_\tau$. This happens iff the interface perpendicular to ${\bf n}$ is stable.}

We can rewrite \eqref{wulff} as
\begin{equation}\label{ext}
W_\tau=\{{\bf x}\,{:}\; \langle\,{\bf x}|{\bf n}\,\rangle\leq \tau({\bf n})\,,\;\forall\,{\bf n}\in{\rm ext}W_\tau^*\}\,.
\end{equation}
Only the interfaces perpendicular to ${\bf n}\in {\rm ext}W_\tau^*$ are stable.
For those ${\bf n}$,  $\tau({\bf n})$ is well-defined and can be measured experimentally.  For ${\bf n}\in \partial W^*_\tau$, but not extremal, a flat interface perpendicular to ${\bf n}$ is unstable and is subject to a hill-and-valley formation built on the interfaces perpendicular to the ${\bf m}\in{\rm ext }W^*_\tau$ entering into the extremal decomposition of ${\bf n}$ (see \cite{H1}).

\noindent
{\bf Remark.\,}
For a relation between these deformations of the unstable interfaces
and the phenomenon of thermal faceting, see \cite{H1} and \cite{W} pp.391-393. It should be stressed however that experiments with crystals and interfaces are rarely done {\em at equilibrium} so that the thermal faceting is basically a non-equilibrium phenomenon.

Mathematically, the support function of $W_\tau$ is defined for any ${\bf n}$, and it can be used to extend the definition of $\tau$. Equivalently, by taking the convex hull of ${\rm ext}W_\tau^*$ we can define $\tau$ for all ${\bf n}\in\partial W_\tau^*$. Hence the hypothesis \eqref{abc} done at the beginning
of section \ref{subsection2.3} are not restrictive.

\noindent
{\bf Example.\,} There is a tangent plane to $W_\tau$ at ${\bf x}$ iff there is a unique support plane $\partial H({\bf n})$ containing ${\bf x}$; the outward  normal to $W_\tau$ at ${\bf x}$ is well-defined and equal to ${\bf n}$.
A tangent plane is always an extremal support plane. Indeed, suppose that
${\bf n}=\lambda{\bf n}_1+(1-\lambda){\bf n}_2$ with  $0<\lambda<1$ and
${\bf n}_1, {\bf n}_2\in {\rm ext}W^*_\tau$; suppose that $\partial H({\bf n})$ is the unique support plane at ${\bf x}\in\partial W_\tau$. Then $\tau({\bf n})=\langle\,{\bf x}|{\bf n}\,\rangle$ and
$$
1=\tau({\bf n})=\lambda \langle\,{\bf x}|{\bf n}_1\,\rangle+(1-\lambda)\langle\,{\bf x}|{\bf n}_2\,\rangle\leq \lambda\tau({\bf n}_1)+(1-\lambda)\tau({\bf n}_2)=1\,.
$$
Therefore
$$
\lambda\underbrace{(\tau({\bf n}_1)-\langle\,{\bf x}|{\bf n}_1\,\rangle)}_{\geq 0}+(1-\lambda)(\tau({\bf n}_2)-\langle\,{\bf x}|{\bf n}_2\,\rangle)=0\,,
$$
and $\tau({\bf n}_i)=\langle\,{\bf x}|{\bf n}_i\,\rangle$ so that $\partial H({\bf n}_i)$ is a support plane at
${\bf x}$. Hence ${\bf n}={\bf n}_1={\bf n}_2$, the decomposition of ${\bf n}$ is trivial and ${\bf n}$ is an extremal point of $W_\tau^*$. We  have extremal support planes which are not tangent planes  when $W_\tau$ has an edge or a corner.

\subsubsection{Subdifferentials of $\tau$}\label{233}

The {\bf subdifferential of $\tau$ at ${\bf x}$} is the set
$$
\partial\tau({\bf x}):=\{{\bf u}\in{\mathbf E}_3\,{:}\; \tau({\bf y})\geq \tau({\bf x})+\langle\,{\bf u}|{\bf y}-{\bf x}\,\rangle\quad\forall\,{\bf y}\}\,.
$$
It is a closed convex set.
An element of $\partial \tau({\bf x})$ is a {\bf subgradient of $\tau$ at ${\bf x}$}. One can prove  that differentiability of $\tau$ at ${\bf x}$ is equivalent to the uniqueness of subgradient;
the unique subgradient is the gradient of $\tau$ at ${\bf x}$,
that is,
$\partial\tau({\bf x})=\{{\rm grad}\tau({\bf x})\}$ and
$$
{\rm grad}\tau({\bf x})=\left(\frac{\partial\tau({\bf x})}{\partial x_1},\frac{\partial\tau({\bf x})}{\partial x_2},\frac{\partial\tau({\bf x})}{\partial x_3}\right)\,.
$$
For ${\bf x}\in\partial W^*_\tau$,  ${\rm grad}\tau({\bf x})$   is normal to the tangent plane  of the polar set $W^*_\tau$ at ${\bf x}$.
Since $\tau$ is sublinear,
$$
\partial\tau({\bf x})=
\partial\tau(t{\bf x})\quad\forall\, t>0\,.
$$
From the definition of the subdifferential of $\tau$ at ${\bf x}={\bf 0}$ one gets
another characterization of $W_\tau$ ($\tau({\bf 0})=0$)
$$
W_\tau=\partial \tau({\bf 0})\,.
$$
{\em For any ${\bf n}\not=0$, the subdifferential $\partial\tau({\bf n})$ is the subset of the boundary of the equilibrium shape $W_\tau$, which is given by the intersection of $W_\tau$ and the support plane $\partial H({\bf n})$, that is
$$
\partial\tau({\bf n})=W_\tau\cap \partial H({\bf n})\quad \text{when ${\bf n}\not=0$}\,.
$$
Moreover, ${\bf v}\in \partial \tau({\bf n})$ iff $\partial{\bf v}^*$ is a support plane of $W^*_{\tau}$ at
${\bf n}/\tau({\bf n})$}.

Indeed, suppose that ${\bf v}\in \partial\tau({\bf n})$. Then
$$
\tau({\bf u})\geq \tau({\bf n})+\langle\,{\bf v}|{\bf u}-{\bf n}\,\rangle\quad\forall\,{\bf u}\,.
$$
Choosing ${\bf u}=0$, we get $\langle\,{\bf v}|{\bf n}\,\rangle\geq \tau({\bf n})$. Choosing ${\bf u}=2{\bf n}$ and using the homogeneity of $\tau$, we get $\tau({\bf n})\geq \langle\,{\bf v}|{\bf n}\,\rangle$. Therefore
$\tau({\bf n})=\langle\,{\bf v}|{\bf n}\,\rangle$ and ${\bf v}\in\partial H({\bf n})$. But this also implies
$$
\tau({\bf u})\geq \tau({\bf n})+\langle\,{\bf v}|{\bf u}-{\bf n}\,\rangle=\langle\,{\bf v}|{\bf u}\,\rangle\quad\forall\,{\bf u}\,,
$$
so that ${\bf v}\in W_\tau$. Conversely, if $\tau({\bf n})=\langle{\bf v}|{\bf n}\,\rangle$ and
$\tau({\bf u})\geq \langle\,{\bf v}|{\bf u}\,\rangle$ for all ${\bf u}$, then
$$
\tau({\bf u})\geq \langle{\bf v}|{\bf u}\,\rangle=\tau({\bf n})+\langle\,{\bf v}|{\bf u}-{\bf n}\,\rangle\quad\forall\,{\bf u}\,.
$$
For the second part notice that if ${\bf v}\in \partial\tau({\bf n})$, then
$\tau({\bf n})=\langle\,{\bf v}|{\bf n}\,\rangle$, so that $\partial{\bf v}^*$ is a support plane of $W_{\tau}^*$ at ${\bf n}/\tau({\bf n})$. Conversely, if $\partial{\bf v}^*$ is a support plane of $W^*_{\tau}$ at ${\bf n}/\tau({\bf n})$, then ${\bf v}\in\partial H({\bf n})$ and
$\langle\,{\bf v}|{\bf u}\,\rangle\leq 1$ for all ${\bf u}\in W_{\tau}^*$, so that ${\bf v}\in W_{\tau}^{**}=W_{\tau}$.

In summary, a corner of $W^*_\tau$ at ${\bf x}$ corresponds to a facet of $W_\tau$ perpendicular to ${\bf x}$ since the subdifferential at ${\bf x}$ is two-dimensional. Edges of $W^*_\tau$ correspond to edges of $W_\tau$. A facet of
$W^*_\tau$ corresponds to a corner of $W_\tau$ since for each ${\bf x}$ of the facet of
$W^*_\tau$ ${\rm grad}\tau({\bf x})$ is the same  and each point of the facet labels a different support plane at ${\rm grad}\tau({\bf x})$.
For the other points ${\bf x}\in\partial W_\tau^*$ the gradient exists and $\langle\,{\rm grad}\tau({\bf x})|{\bf x}\,\rangle=1$;
at ${\rm grad}\tau({\bf x})\in W_\tau$ there is a unique (extremal) support plane perpendicular to ${\bf x}\in W^*_\tau$.

\bigskip

\noindent
{\bf Example.\,}
For the Ising model, at zero temperature, $W_\tau=\{{\bf x}\,{:}\; \|{\bf x}\|_\infty\leq 1\}$ and
$W_\tau^*=\{{\bf x}\,{:}\; \|{\bf x}\|_1\leq 1\}$. In dimension $d$ there are only $2d$ extremal points
for $W_\tau^*$ corresponding to the $2d$ facets of $W_\tau$. All other interfaces  are unstable.
For $d=2$, at non-zero temperature, the corners of  $W_\tau^*$ are smoothed out, so that $W_\tau$ has no facet. Moreover, $W_\tau^*$ is strictly convex, so that $W_\tau$ has no corner and all points of $\partial W_\tau^*$ are extremal points, hence all interfaces are stable. The disappearing of facets is called roughening transition; here the temperature of the roughening transition is zero. In the rough phase, i.e. above zero temperature, ${\bf \tau}$ is differentiable. This situation is generic for two-dimensional cases at non-zero temperature. For $d=3$, if the temperature is strictly positive and  strictly smaller than the two-dimensional critical temperature, inequalities \eqref{step} and \eqref{comparison} imply that
there are facets for $W_\tau$ corresponding to corners for $W_\tau^*$.
Notice that by symmetry, instead of taking ${\bf n}(\theta)=(0,-\sin\theta,\cos\theta)$ as in subsection \ref{213}, we may also choose ${\bf n}(\theta)=(-\sin\theta,0,\cos\theta)$ or
${\bf n}(\theta)=(-\sin\theta,\cos\theta,0)$.
At a higher temperature the system undergoes a roughening transition with the disappearing of facets, but this result has not yet been mathematically established for the Ising model.

\subsubsection{Radius of curvature and stability  of interfaces}\label{234}

In this subsection we assume that $\tau$ is strictly convex and smooth (except at the origin). At each point of $\partial W_\tau$ there is a  well-defined normal ${\bf n}$ and
$W_\tau\cap \partial H({\bf n})=\{{\rm grad}\tau({\bf n})\}$.
We consider the two-dimensional case  $W_\tau\subset {\mathbf E}_2$.
We set for $\alpha\in [0,2\pi)$,
$$
{\bf n}(\alpha):=(\cos\alpha,\sin\alpha)\quad\text{and}\quad{\bf m}(\alpha):=(-\sin\alpha,\cos\alpha)\,.
$$
The vectors ${\bf n}$ and ${\bf m}$ are orthonormal and we can decompose ${\rm grad}\tau({\bf n})$ as
$$
{\rm grad}\tau({\bf n})=\langle\,{\rm grad}\tau({\bf n})|{\bf n}\,\rangle{\bf n}+
\langle\,{\rm grad}\tau({\bf n})|{\bf m}\,\rangle {\bf m}=
\tau({\bf n}){\bf n}+ \langle\,{\rm grad}\tau({\bf n})|{\bf m}\,\rangle {\bf m}\,,
$$
since $\tau({\bf n})=\langle\,{\rm grad}\tau({\bf n})|{\bf n}\,\rangle$. We also have
$$
\frac{d}{d\alpha}\tau({\bf n}(\alpha))=
\langle\,{\rm grad}\tau({\bf n}(\alpha))|{\bf m}(\alpha)\,\rangle\,.
$$
Therefore we can parametrize $\partial W_\tau$ by
$$
\alpha\mapsto {\rm grad}\tau({\bf n}(\alpha))=\tau({\bf n}(\alpha))\,{\bf n}(\alpha)+
\frac{d}{d\alpha}\tau({\bf n}(\alpha))\,{\bf m}(\alpha)\,,\quad\alpha\in[0,2\pi)\,.
$$
From this expression one obtains
$$
\frac{d}{d\alpha}{\rm grad}\tau({\bf n}(\alpha))=\left(\tau({\bf n}(\alpha))+
\frac{d^2}{d\alpha^2}\tau({\bf n}(\alpha))\right){\bf m}(\alpha)\,,
$$
and  the radius of curvature $\rho(\alpha)$ at ${\rm grad}\tau({\bf n}(\alpha))$, which is given by
\begin{equation}\label{radius}
\rho(\alpha)=\tau({\bf n}(\alpha))+
\frac{d^2}{d\alpha^2}\tau({\bf n}(\alpha))\,.
\end{equation}
Strict positivity of $\rho(\alpha)$ means absence of corner at ${\rm grad}\tau({\bf n}(\alpha))$.
There is an interesting inequality, noticed by Ioffe, which strengthens the stability condition \eqref{stable}. {\em If the radius of curvature $\rho(\alpha)$ is  bounded below uniformly,
$$
\inf_\alpha \rho(\alpha)=\chi>0\,,
$$
then}
\begin{equation}\label{strongineq}
\tau({\bf x}) +\tau({\bf y})-\tau({\bf x}+{\bf y})\geq \chi(\|{\bf x}\|+\|{\bf y}\|-\|{\bf x}+{\bf y}\|)\quad\forall\,{\bf x},{\bf y}\,.
\end{equation}
The constant $\chi$ is the best possible constant.

In the three-dimensional case, at ${\rm grad}\tau({\bf n})\in \partial W_\tau$,
${\bf n}$ is  normal to $\partial W_\tau$.
To study the curvature at this point one can slice $W_\tau$ by planes containing ${\bf n}$. This reduces the problem to two-dimensional situations.  If the radius of curvature in each slice is bounded below uniformly by $\chi>0$, then  we get a stronger version of stability inequality \eqref{2.3}: for any tetrahedron ${\mathcal T}(A_0,A_1,A_3,A_3)$
$$
|\Delta_1|\tau({\bf n}_1)+|\Delta_2|\tau({\bf n}_2)+|\Delta_3|\tau({\bf n}_3)-|\Delta_0|\tau({\bf n})\geq \chi (|\Delta_1|+|\Delta_2|+|\Delta_3|-|\Delta_0|)\,.
$$
Indeed, for ${\bf x}=|\Delta_1|{\bf n}_1$, ${\bf y}=|\Delta_2|{\bf n}_2$ and ${\bf z}=|\Delta_3|{\bf n}_3$
$$
\tau({\bf x})+\tau({\bf y}+{\bf z})-\tau({\bf x}+{\bf y}+{\bf z})\geq\chi(\|{\bf x}\|+\|{\bf y}+{\bf z}\|-\|{\bf x}+{\bf y}+{\bf z}\|)
$$
and
$$
\tau({\bf y})+\tau({\bf z})-\tau({\bf y}+{\bf z})\geq\chi(\|{\bf y}\|+\|{\bf z}\|-\|{\bf y}+{\bf z}\|)\,.
$$

\subsubsection{Isoperimetric inequality}\label{235}

Gibbs (1878) and Curie (1885)  studied a special case of the following variational problem concerning the equilibrium shape of a crystal, and Wulff (1901) gave the geometrical interpretation of the solution showing that the equilibrium shape is obtained by the construction leading to \eqref{wulff}. For this reason
the equilibrium shape $W_\tau$ is often called {\bf Wulff crystal}.

Suppose that $V\subset {\mathbf R}^d$, $d=3$ or $2$, and that ${\bf n}(s)$ denotes the outward unit normal to its boundary $\partial V$ at $s$. The surface free energy of this set is given by the surface integral
$$
{\mathcal F}_\tau(\partial V):=\int_{\partial V}\tau({\bf n}(s))\,d{\mathcal H}^{d-1}(s)\,.
$$
($d{\mathcal H}^{d-1}$ is the $(d-1)$-Hausdorff measure in ${\mathbf R}^d$.)
The problem considered by Wulff was to determine the set $V$, which minimizes the functional ${\mathcal F}_\tau(\partial V)$ among a class of subsets with  fixed volume. We state an isoperimetric inequality which gives the solution to this problem. Roughly speaking, the  subsets which can be considered are those subsets for which the functional ${\mathcal F}_\tau$ is well-defined. Denote by $|C|$ the volume of the subset $C\subset {\mathbf R}^d$.
Then
\begin{equation}\label{isoperimetric}
{\mathcal F}_\tau(\partial V)\geq d|W_\tau|^{1/d} |V|^{(d-1)/d}\,.
\end{equation}
Equality holds if and only iff $V=W_\tau$ up to dilation and translation.

\subsection{Summary}\label{summary}

The subject of this article is the definition of the interface free energy $\tau$
and its thermodynamical properties once the interatomic interactions of the system are given.  Provided that one can construct a macroscopic state with a planar interface perpendicular to ${\bf n}$,  one can use formula \eqref{1.3} to obtain $\tau({\bf n})$.
The fundamental property of the interface free energy is that it is a convex function.
The interface free energy can be measured experimentally at equilibrium only for the interfaces
which are thermodynamically stable.  By convention the physical value of the interface free energy $\tau({\bf n})$ is given for
a unit vector ${\bf n}$. But, using the extension of $\tau$ as an homogeneous function, this function can be interpreted either as the support function of the equilibrium shape
$W_\tau=\{{\bf x}\,{:}\; \langle\,{\bf x}|{\bf n}\,\rangle\leq \tau({\bf n})\,,\;\forall\, {\bf n}\}$,
or as the gauge function of
$W^*_\tau=\{{\bf x}\,{:}\; \tau({\bf x})\leq 1\}$.
Stable interfaces are labeled by the extremal points of $W^*_\tau$.

The interface free energy is considered here only from the macroscopic viewpoint; for that reason
the macroscopic states of are the relevant states. These states should not be confused with the infinite-volume Gibbs measures, which describe the states of the system at the microscopic scale, when one chooses as reference length the lattice spacing. For example, for the two-dimensional Ising model one has macroscopic states with stable interfaces for all directions ${\bf n}$, while there are only translation invariant infinite-volume Gibbs measures. The study of interfaces at the microscopic scale, when the interfaces are stochastic geometrical objects, is of course an important, related but different topic (see references for section \ref{section2}).
When one studies interfaces at the microscopic scale, it is often natural and more convenient to  replace the definition of the interface free energy by another definition, which is based on specific microscopic properties of the system. One essential point is to prove that the definition used is the same as the definition \eqref{1.3}.

\section{Bibliographical Notes}\label{section4}

\noindent
{\bf Section \ref{section2}.\,} \cite{H2}, \cite{RW2}, \cite{W}, \cite{Z} are reviews of physics on interfaces and equilibrium shapes of crystals. The reviews  \cite{A} and \cite{P1} are more specifically concentrated on the interface free energy; they are reviews of mathematical physics. The excellent review
\cite{A} also contains  a lot of information on related subjects. Some results concerning the study of interfaces at the microscopic scale are discussed in \cite{P1}.
The  book \cite{Pr} contains a lot of  material related to the interface free energy, which is not presented here.

The problem of the spatial distribution of the coexisting phases, starting with the monograph
\cite{DKS}, and the papers \cite{ACC} and \cite{P2},
has been one of the major research themes in mathematical statistical mechanics during the last decade of the XXth century, for which deep mathematical results have been obtained for several models.
The interface free energy (as defined in section \ref{212}) enters in an essential way in the formulation of the results, and interfaces are studied at the microscopic scale.
See the review paper \cite{BIV}, the mathematical monograph \cite{C} and \cite{Pr}.

\noindent
{\bf Section \ref{211}.\,}
In relation with the topic of the article, chapter 6 of \cite{G} is  a good introduction to the Ising model, but contains few recent results. An up-to-date  good reference is \cite{V}.

\noindent
{\bf Section \ref{212}.\,}
The macroscopic limit for the two-dimensional Ising model is discussed in \cite{PV}. Using the methods exposed in \cite{BIV} analogous results can be obtain for the three-dimensional case. The up-to-date
reference concerning proofs of existence and convexity of surface tension for ferromagnetic models is \cite{MMR}. For more general models see  \cite{BP}.
The role of the complete wetting of the walls in the definition \eqref{tau} has been explicitly  emphasized  in \cite{PV}.
The physics of a situation where the interface is partially pinned to the wall had  been studied  for the two-dimensional Ising model by Abraham and Ko already in \cite{AK}.
Mathematical results on wetting phenomenon for Ising systems are in \cite{FP} and \cite{PV2}. Wall free energies are treated in \cite{FC}.

Comparisons of several definitions of the interface free energy are carefully discussed in \cite{A} and references can be found there.
The computation of Onsager is in his famous paper where he solved two-dimensional Ising model at zero magnetic field \cite{O}; the computation of Abraham and Martin-L\"of is in \cite{AM}. The computation of the full interface free energy is detailed in the treatise of McCoy and Wu \cite{MW}. A reference where the results of the computation can be easily found is \cite{RW1}.

\noindent
{\bf Section \ref{213}.\,}
Inequality \eqref{above} is proven in  \cite{BLP2} and \eqref{below} in \cite{LP}.
Step free energy is discussed in \cite{A} and in \cite{BEF}. Inequality \eqref{step} is proven in \cite{BEF} and inequality \eqref{comparison} in \cite{BFL}.

\noindent
{\bf Section \ref{subsection2.2}.\,} The basic reference is \cite{H1}.

\noindent
{\bf Section \ref{subsection2.3}.\,} The basic reference is \cite{H1}, but Herring uses the {\bf surface tension plot}, which is the set of points
$$
\{{\bf x}\in {\mathbf E}_3\,{:}\; {\bf x}=\tau({\bf n})\,{\bf n}\,,\;\|{\bf n}\|=1\}\,,
$$
for studying $\tau$. This is the standard way of presenting $\tau$ in physics.
One gets the surface tension plot from $\partial W_\tau^*$ by an inversion on the unit sphere (or the unit circle in dimension 2). Affine parts of $\partial W_\tau^*$ become spherical parts, or circular parts, of the surface tension plot. Using the surface tension plot
the fundamental convexity property of $\tau$ is hidden (this property  is not stated explicitly in \cite{H1});
one also looses the natural labeling of the stable interfaces by the extremal points of $W_\tau^*$ of subsection \ref{232}. The duality between non-zero ${\bf x}$ and half-spaces ${\bf x}^*$ is the natural language for
discussing the geometry of the equilibrium shape. The use of the polar
set $W_\tau^*$ as an alternative to the surface tension plot is mentioned in \cite{MMR}.
The surface tension plot of the two-dimensional Ising model is analyzed   in \cite{RW1}.

Inequality \eqref{strongineq} appears in \cite{I1} in a slightly different form; the formulation given  is taken from \cite{PV}.

The mathematics of subsections \ref{231} to \ref{234}  is classical and  treated thoroughly  in \cite{M}. An accessible modern reference is chapter 1 of \cite{S}.

The isoperimetric inequality is  a classical topic in analysis. There are many papers on  inequality \eqref{isoperimetric}, e.g. \cite{D} (one of the early proofs),  \cite{T1} and \cite{F}. The proof of \eqref{isoperimetric} in  these papers is based on Brunn-Minkowski inequality. In dimension two a different approach is possible \cite{DP}.

\end{document}